\documentclass[a4paper]{jpconf}
\usepackage{graphicx}
\usepackage[dvips]{color}
\usepackage{inputenc}

{}
\newcommand{\BAN}{\ensuremath{B_{1g}\,}}
\newcommand{\BN}{\ensuremath{B_{2g}\,}}

\newcommand{\OmN}{\ensuremath{\Omega_{B_{2g}}\,}}
\newcommand\Deltam{\ensuremath{\Delta_{max}\,}}
\newcommand{\ZAN}{\ensuremath{Z\Lambda_{AN}\,}}
\newcommand{\ZN}{\ensuremath{Z\Lambda_{N}\,}}
\newcommand{\vD}{\ensuremath{v_\Delta\,}}

\newcommand{\arc}{\ensuremath{f_c\,}}

\begin{document}
\title{New insights into the phase diagram of the copper oxide superconductors from electronic Raman scattering}

\author{A. Sacuto, Y. Gallais, M. Cazayous, M.-A. M\'easson}

\address{Laboratoire Mat\'eriaux et Ph\'enom$\grave{e}$nes Quantiques (UMR 7162 CNRS),
Universit\'e Paris Diderot-Paris 7, Bat. Condorcet, 75205 Paris Cedex 13, France}

\author{G. D. Gu}
\address {Matter Physics and Materials Science, Brookhaven National Laboratory (BNL), Upton, NY 11973, USA}

\author{D. Colson}
\address {Service de Physique de l'Etat Condens\'e, CEA-Saclay, 91191 Gif-sur-Yvette, France}

\ead{alain.sacuto@univ-paris-diderot.fr}

\begin{abstract}
Mechanism of unconventional superconductivity is still unknown even if more than 25 years have been passed since the discovery of high-Tc cuprate superconductors by J.G. Bednorz and K. A. Muller \cite{Bednorz}. Here, we explore the cuprate phase diagram by electronic Raman spectroscopy and shed light on the superconducting state in hole doped cuprates. Namely, how superconductivity and the critical temperature $T_c$ are impacted by the pseudogap.  
\end{abstract}

\section{Introduction}

In conventional superconductors described by BCS theory \cite{BCS}, superconductivity emerges from a metal and the whole of electronic states around the Fermi surface condense into the superconducting state. In hole doped cuprates, the situation is more subtle, a 
metal is built by holes doping into a Mott insulator state \cite{Anderson,Kotliar,Lee}. Superconductivity emerges in a finite doping range and 
the k-space distribution of the electronic states involved in superconducting state is strongly doping dependent. 
At low doping level, most of electronic states involved in superconducting state are localized around the diagonal of the Brillouin zone (nodal region). 


Such a change in the electronic distribution around the Fermi surface as a function of doping level is the missing link which a long time have disturbed physicists for understanding the cuprate phase diagram \cite{Norman,Millis,Hufner,Basov}. This change likely comes from the emergence of a competing order with superconductivity, the pseudogap phase \cite{Timusk} which reduces the electronic states available for superconductivity around the principal axes of the Brillouin zone (antinodal region). 

Here, we argue from an electronic Raman scattering study that the superconducting dome circumscribed by $T_c$ is divided in three regions (see fig.1): a high doping level region  where superconductivity develops on the entire Fermi surface,  an intermediate doping level region where superconductivity is weakened at the antinodes and a low doping level region where superconductvity is essentially confined around the nodes. For these two last parts, the critical temperature $T_c$ is no more controled solely by the superconducting gap. The fraction of coherent Fermi surface where superconductivity settles has to be considered. This gives rise to the emergence of a new relationship between the superconducting gap and $T_c$ (distinct from the BCS one available for conventional superconductor) and reveals the compromise which has to be found for increasing $T_c$. 
 

\begin{figure}[!ht]
\begin{center}
\includegraphics[width=12cm]{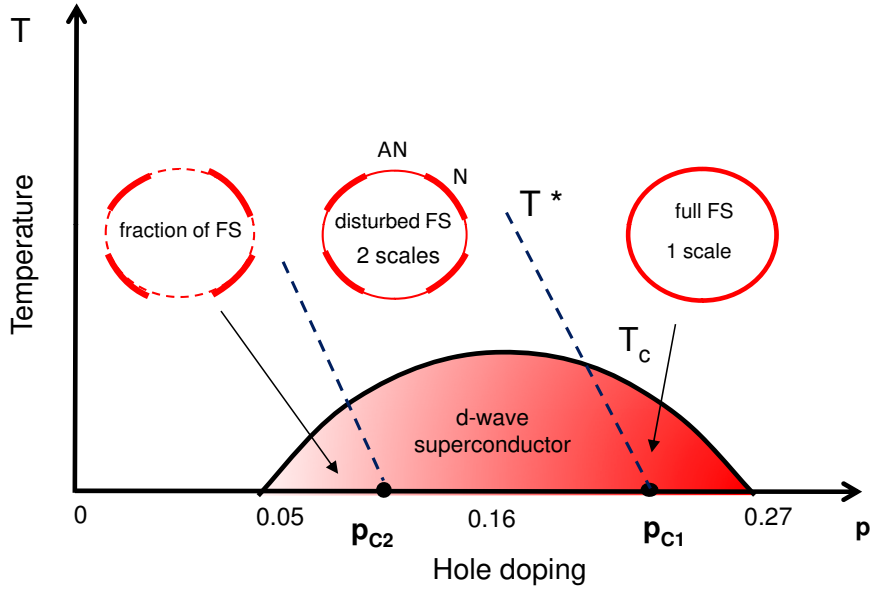}
\end{center}\vspace{-7mm}
\caption{The superconducting dome is divided in three parts depending on the loss of coherent superconducting excitations (coherent Bogoliubov quasiparticles) along the principal axes of the Brillouin zone (antinodal region, AN): a full Fermi surface at high doping level, a disturbed Fermi surface for intermediate doping level (including the optimal one for where $T_c$ reaches its maximum) 
and a fractionalized Fermi surface at low doping level.}
\label{fig1}
\end{figure}

\section{Two critical doping levels revealed from the exploration of the cuprate phase diagram by electronic Raman scattering}

Identifying the changes in the electronic properties inside the superconducting dome is the first objective. 
In figure 2 is shown the temperature dependence of the Raman spectra of Bi-2212 ($Bi_2Sr_2CaCu_2O_{8+\delta}$) single crystals for several doping levels starting from $p=0.22$ to $0.10$. Temperatures are ranging from well below $T_c$ to approximatively $10~K$ above $T_c$. The first and second panels exhibit the Raman spectra in $B_{1g}$ and $B_{2g}$ geometries which correspond respectively to the principal axes and the diagonal of the Brillouin zone (BZ). In these geometries, we probe the antinodal (AN) and nodal (N) regions where the amplitude of a $d-$wave superconducting gap is expected to be maximum and vanished respectively \cite{Devereaux}. We observe in these both geometries a broad peak which gradually decreases in intensity as the crystal is heated up to $T_c$ for disapearing at $T_c$. 
These peaks are 
assigned to coherent peaks of the superconducting state as demonstrated in earlier investigations \cite{Blanc}. 
The low energy \BN Raman response exibits a linear frequency dependence in the superconducting state. This signals the existence of nodes in the superconducting gap along the diagonal of the BZ.  On the contrary the low energy \BAN response is roughtly cubic in frequency which signals a maximum amplitude for the superconducting gap along the principal axes of the BZ\cite{Devereaux}. 

\begin{figure}[!ht]
\begin{center}
\includegraphics[width=16cm]{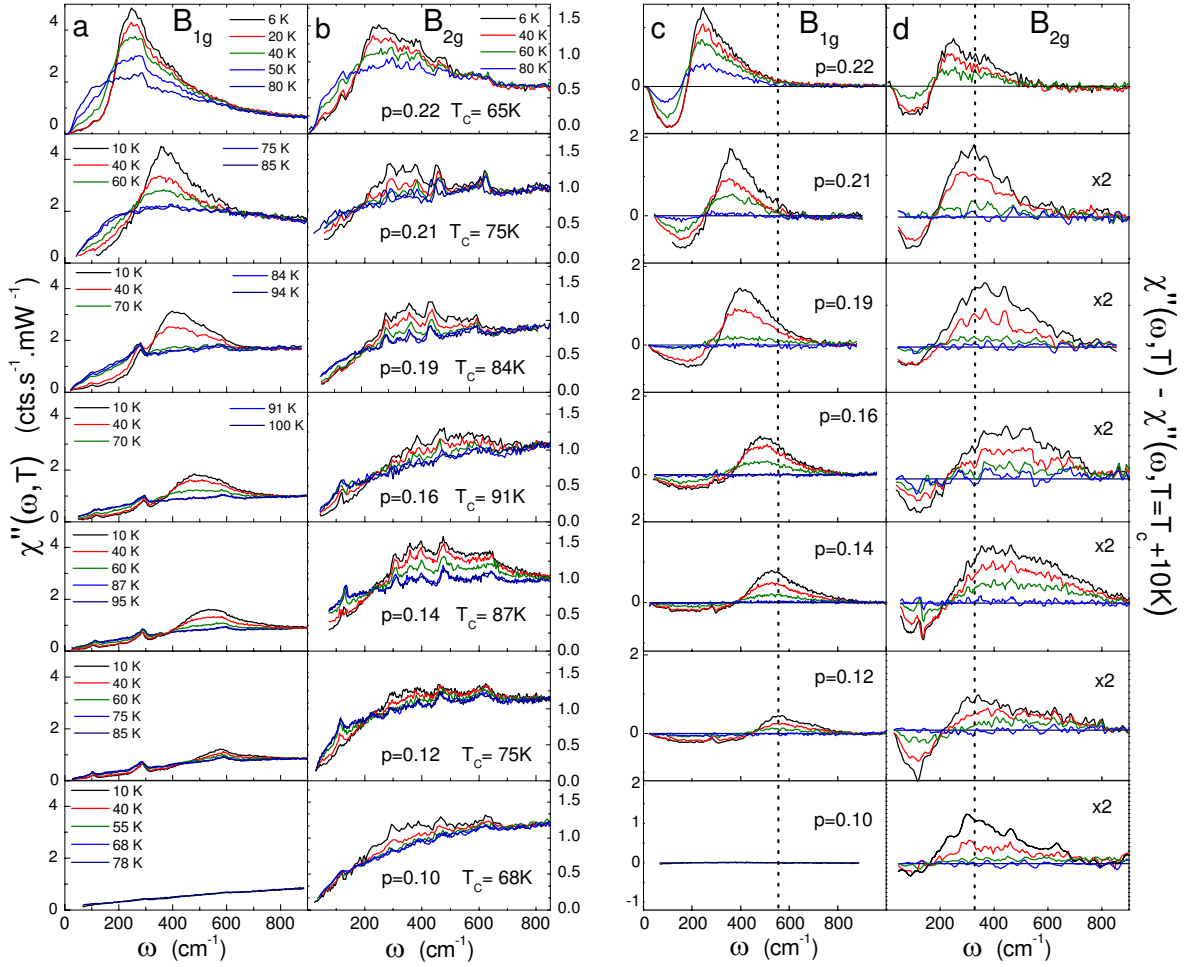}
\end{center}\vspace{-7mm}
\caption{Temperature dependence of the Raman spectra of Bi-2212 for several doping levels (from overdoped ($p>0.16$) to underdoped ($p<0.16$)) in (a) \BAN and (b) \BN geometry. In these two first panels,  special care has been devoted to make reliable quantitative comparisons between the Raman intensities of distinct crystals with different doping levels measured in the same geometry, and between measurements in distinct geometries for crystals with the same doping level. Raman spectra subtracted from the one measured at $10~K$ above $T_c$ in (c) \BAN and (d) \BN geometry. In the last panel, the \BN Raman intensities (excepted for $p=0.22$) have been magnified by a factor of 2 in order to emphasize the temperature evolution of the \BAN and \BN peaks.\cite{Blanc}}
\label{fig2}
\end{figure}

Two experimental observations can be pointed out from these spectra. First, the doping dependences of the \BAN and \BN peaks energy are distinct (see panels third and fourth of fig.2). The energy of the \BAN peak increases with underdoping while the energy of the \BN peaks increases up to the optimal doping $p=0.16$ before decreasing. Second, the intensity of the \BAN peak is strongly affected as the doping level falls, it decreases before disappearing close to $p=0.10$ (see first panel of fig.2). On the contrary, the intensity of the \BN peak remains sizeable even at low doping level (see second panel of fig.2).

We focus on the first observation. In figure 3-a is reported the doping evolution of \BAN and \BN peak positions for Bi-2212 and Hg-1201 compounds. At high doping level the \BAN and \BN  peak energies exhibit the same doping dependence : they increase as the doping level decreases down to approximatively $p=0.20$. Below this doping level two energy scales appear \cite{leTacon,Opel}. The \BAN energy scale increases almost linearly with underdoping while the \BN one follows $T_c$ ($6~k_BT_c$): it increases up to the maximum of $T_c$ and then decreases. We can define experimentally a critical doping level namely, $p_{c1}=0.20 $, where the two energy scales appear in the superconducting state. 

Our second experimental observation is the collapse of the \BAN peak intensity as the doping level is reduced while the intensity of the \BN peak remains sizeable \cite{Blanc2}. 
In figure 3-b we have reported the normalized integrated area under the \BAN and \BN peaks. (They are obtained from the subtraction between the Raman reponses well below $T_c$ and just above $T_c$). We see that the integrated area of the \BAN peak (full square) decreases as the doping level is reduced and vanishes for a doping level close to $p=0.10$. This defines a second critical doping level denoted $p_{c2}$ close to $p=0.10$. $p_{c1}$ and $p_{c2}$ delimit the boundaries of the intermediate region (see fig. 1).

\begin{figure}[!ht]
\begin{center}
\includegraphics[width=12cm]{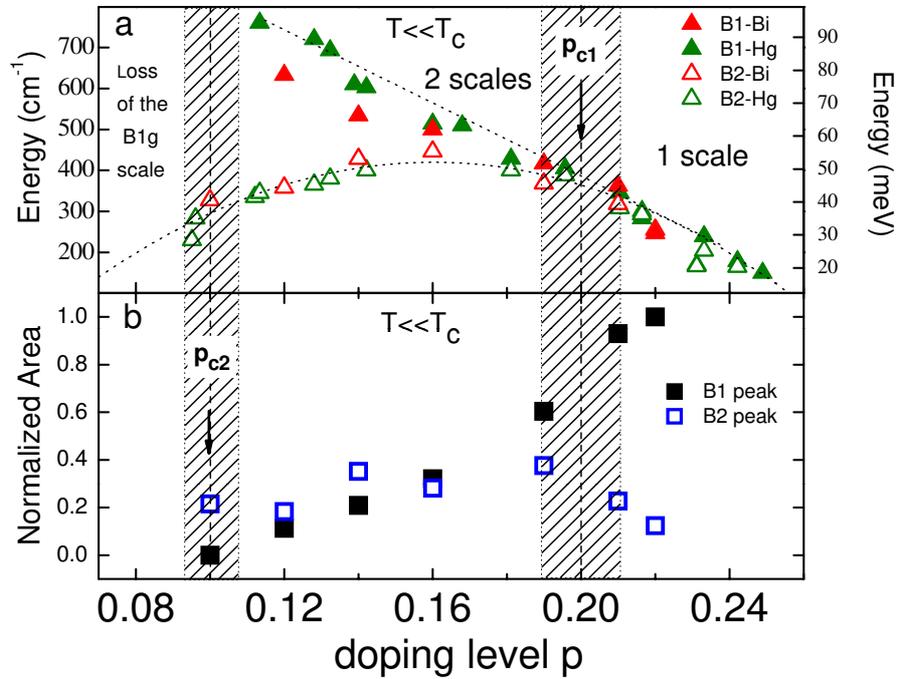}
\end{center}\vspace{-7mm}
\caption{(a) Doping evolution of the \BAN and \BN peak positions extracted from the Raman spectra of Bi-2212 and Hg-1201 compounds \cite{Blanc,leTacon}. Around $p_{c1}=0.20$, two energy scales appear. (b) Doping evolution of the integrated area under the \BAN and \BN superconducting peaks extracted from the Raman spectra of Bi-2212. Around $p_{c2}=0.10$, the intensity of  the \BAN peak collapses. On the opposite the area of the \BN peak (open square) doesn't vanish even at low doping level. All the data have been measured in the superconducting state well below $T_c$.}
\label{fig3}
\end{figure}

The \BAN and \BN peak area can provide a direct estimate of the density of Cooper pairs in the nodal and antinodal regions. Indeed for a non interacting Fermi liquid, in the framework of BCS theory, we can show that the integral of the Raman response over $\Omega$ when at $T=0~K$, gives \cite{Blanc2} : 

\begin{eqnarray}          
\int{\chi^{,,}_{\mu}(\Omega)d\Omega}=4\pi  \sum_{k}(\gamma^{\mu}_{k})^{2}\sum_{k}(u_kv_k)^2
 \label{eq6}
\end{eqnarray}           
 where $\mu $ refers to the $B_{1g}$ and $B_{2g}$ geometries, $\gamma^{\mu}_{k}$ is the Raman vertex, $v_k^2$ and $u_k^2$ are the probabilities of the pair $(k\uparrow,-k\downarrow)$ being occupied and unoccupied respectively. This sum is non-vanishing only around the Fermi energy $E_F$ in the range of the superconducting gap $2\Delta_k$ \cite{deGennes}. This quantity corresponds to the density of Cooper pairs, formed around the Fermi level as the gap is opening \cite{Leggett}. 
The integral of the Raman response is then proportional to the density of Cooper pairs, weighted by the square of the Raman vertex which selects specific parts of the Brillouin zone: the nodal or the antinodal regions. In summary, the Raman response function exhibits the coherent Bogoliubov quasiparticles which are the excitated states of the Cooper pairs and its integral over $\Omega$ gives a direct estimate of the density of Cooper pairs. 

Applying this analysis to our data reveals that at high doping level, the density of Cooper pairs is significant in both the nodal and antinodal region. However it decreases at the antinodes as the doping level is reduced and vanishes close to $p_{c2}=0.10$, while it is still sizeable around the nodes. Therefore we are led to conclude that the density of Cooper pairs is strongly doping dependent and for low doping level it is confined in k-space. Cooper pairs are then forming k-space islands around the nodes. Clues of a such k-space electronic distribution have been recently suggested by tunneling and photoemission measurements where loss of coherent Bogoliubov quasiparticles at the antinodes has been reported ~\cite{McElroy,Shen,Kohsaka,Ding,Feng,Kondo,Vishik}. Since spectroscopy techniques detect coherent Bogoliubov quasiparticles which are the excitated states of Cooper pairs,  
these measurements are consistent with theoretical pictures where most of the supercurrent is carried out by electrons small patches centered on the nodal points \cite{Ioffe} and the robustness of nodal fermions \cite{Tesanovic}. The doping evolution of the density of Cooper pairs is sketched in fig.~4. Although it seems counter-intuitive, we find that superconductivity is robust in the nodal region where the amplitude of the superconducting gap is the smallest.

\begin{figure}[!ht]
\begin{center}
\includegraphics[width=12cm]{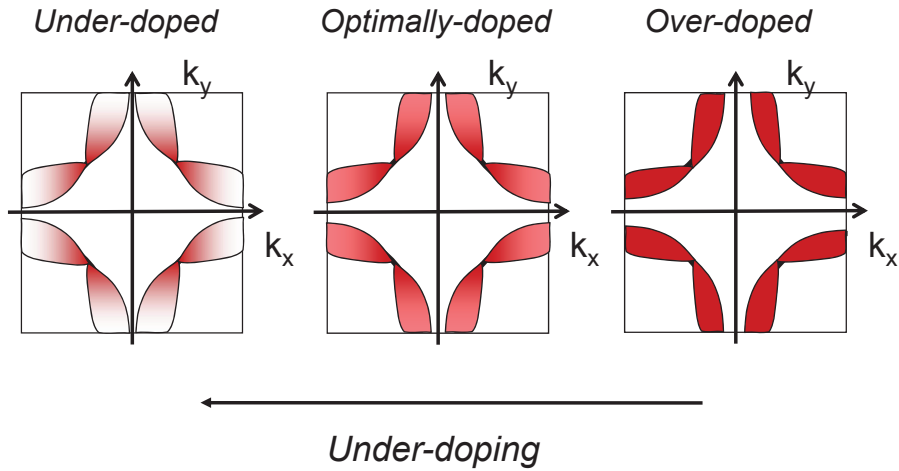}
\end{center}\vspace{-7mm}
\caption{Sketches of the d-wave superconducting gap amplitude in the momentum space for three distinct doping levels. The dark (red) zone corresponds to a high density of coherent Cooper pairs and the bright one to a low density of Cooper pairs. Cooper pairs develop preferentially around the diagonal of the Brillouin zone with falling doping and form like k-space islands of Cooper pairs.}
\label{fig4}
\end{figure}

\section{Impact of the pseudo gap on the superconducting state}

In order to get a better understanding of the cuprate phase diagram we have to determine the origin of the emergence of the two energy scales at the $p_{c1}$ doping level and the strong loss of Cooper pairs density in the antinodal region at $p_{c2}$,  giving rise to the confinement of Cooper pairs in the nodal region. For this, we have to consider the pseudogap phase which has been originally discovered by nuclear magnetic resonance (NMR) \cite{Alloul,Warren} and then extensively studied by transport \cite{Hussey} and spectroscopic probes \cite{Damascelli,Fischer,Timusk}.

The pseudogap can be defined as a loss of electronic states around the Fermi level which are only restored above the pseudogap temperature $T^*$. Its effect on low energy electronic states is mainly in the antinodal region. 
The pseudogap manifests itself in the \BAN Raman response function by a low energy depletion of the electronic background as the temperature decreases \cite{Opel,Blumberg,Nemetschek,Gallais,Guyard,Sacuto}. In the left pannel of fig. 5 is displayed the temperature dependence of the Raman spectra of Bi-2212 crystals for various doping levels.  The thick curves (red and black) underline the depletion in the low energy range of the Raman spectra. There is no depletion for the overdoped (OD65K) compound. The low energy electronic background increases as the temperature decreases. This is expected for a metal ~\cite{Devereaux}. Except for the OD 65 K,  we can estimate the pseudogap temperature $T^*$ by studying the temperature dependence of the normalized integrated area of the $B_{1g}$ Raman response (up to $800~cm^{-1}$). We have defined $T^*$ as the temperature for which the low energy electronic background intensity is restored.  Once the low energy electronic background is restored, it shows a metallic behaviour and decreases in intensity as the temperature raises. This corresponds to the maximum of the integrated area. The integrated area are plotted in the middle panel of fig. 5 and the $T^*$ values are indicated by dashed lines.  We find $T^* \approx 210,~185$ and $160~ K$ for the underdoped (UD) 75 K and 85 K  and overdoped (OD) 84 K  coumpounds respectively.

\begin{figure}[ht!]
\begin{center}
\includegraphics[width=12cm]{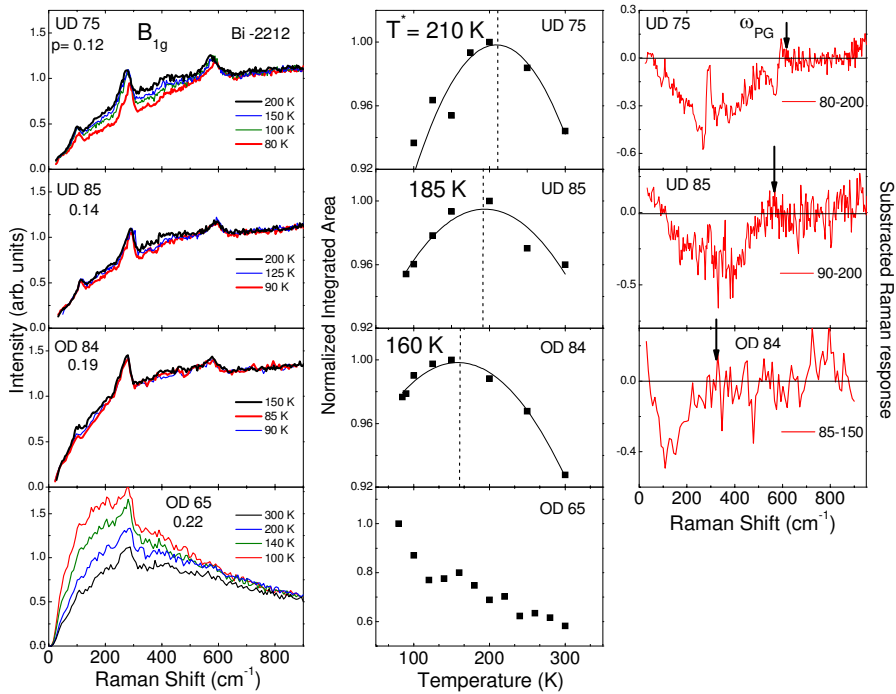}
\end{center}\vspace{-7mm}
\caption{Left panel: selected temperature dependences of the \BAN Raman spectra of Bi-2212 for several doping levels. Thick lines underline the low energy electronic background depletion. Middle panel:  Normalized integrated area for the $B_{1g}$ Raman spectra. Normalization was achieved by dividing the integrated area by its maximum value. The vertical dotted line indicates $T^{*}$ and the thin line corresponds to a polynomial fit. Right panel: subtraction between the Raman spectra measured at  $T^*$ and just above $T_{c}$. No $T^{*}$ and $\omega_{PG}$ values were detected for the OD 65 crystal. Notice that the shape of electronic background changes drastically with doping between OD 84 and OD 65.~\cite{Sacuto}} 
\label{fig5}
\end{figure}


We can also define the pseudogap energy by subtracting the Raman response measured close to $T^*$  from the one just above $T_c$. The subtracted spectra are shown in the right panel of fig. 5. The pseudogap energy $\omega_{PG}$ is defined as the end of the energy range inside which the depletion sets in (see arrow). 

We find that $T^*$ and $\omega_{PG}$ increase with underdoping with $\omega_{PG}\approx 4.3T^{*}$ in good agreement with previous studies on the pseudogap \cite{Kugler}.  


Our study on the pseudogap reveals that the pseudogap emerges between $0.19$ and $0.22$ very close to the $p_{c1}$ doping level. This is illustrated in fig. 6  where the normalized area of the electronic background depletion (fingerprint of the pseudogap in \BAN Raman spectra) is reported as a function of doping level (open star). We clearly see that the depletion opens close to $p_{c1}$. This is the doping range (shaded zone) for where the two energy scales appear in the superconducting state, see fig.3. 

In fig. 6, the area of the \BAN peak varies in opposite manner with the strength of the pseudogap. The area of the \BAN peak decreases for vanishing at $p_{c2}$ while the pseudogap depletion becomes more and more pronounced as the doping level is reduced. As seen previously, the \BAN peak area provides a direct estimate of the density of Cooper pairs 
in the antinodal region.
This gives us experimental evidence that the pseudogap acts again the formation of Cooper pairs in the antinodal region. 

It has not escaped to the reader that $p_{c2}$ is a key doping level several times mentioned by many experimental studies: neutrons \cite{Tranquada}, scanning tuneling spectroscopy (STS) \cite{McElroy,Wise}, NMR \cite{Wu}, transport properties \cite{Sun,Daou} and resonant soft X-rays scattering \cite{Ghiringhelli}. 
 
At this specific doping level fluctuating charge-density wave or magnetic field induced charge density wave or combination of charge and spin orders have been reported in hole doped cuprates. $p_{c2}$  seems to be the specific point for which 
an effective magnetic field induced rearrangement of the Fermi surface is observable \cite{Doiron,leBoeuf,Laliberte}. 
Focusing on Bi-2212 compound near $p_{c2}$ a checkerboard charge ordering have been detected from STS (without magnetic field) which is coincident with antinodal quasiparticle decoherence \cite{McElroy}. 

In our interpretation the superconducting gap and the pseudogap coexist and compete each other below $p_{c1}$. 
This is also supported by earlier and recent studies \cite{Loram,Tallon,Bernhard,Fauque, Kohsaka2,Taillefer,He,Florence,Sakai}. 


\begin{figure}[ht!]
\begin{center}
\includegraphics[width=10cm]{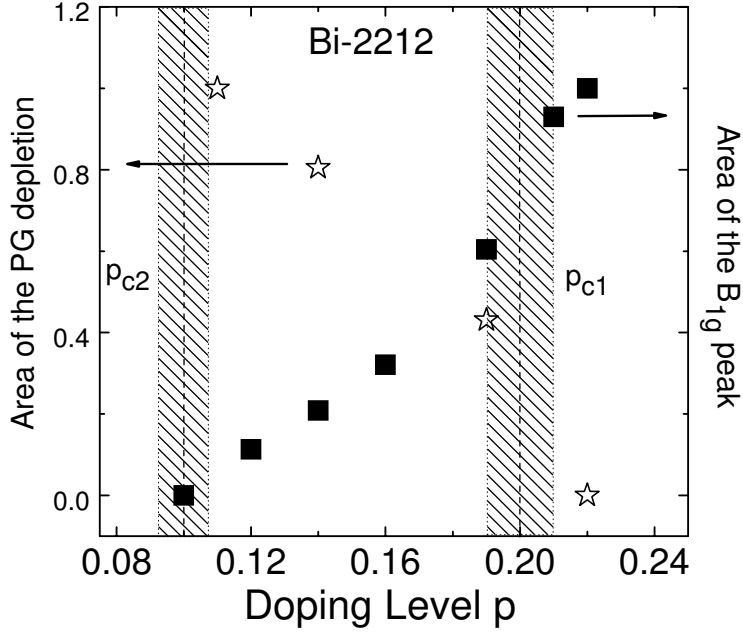}
\end{center}\vspace{-7mm}
\caption{ Doping evolution of the normalized area of the pseudogap depletion and the pair breaking peak in \BAN geometry. Normalization was achieved from dividing the area by their maximum values in the doping range of interest.
 Notice that the pseudogap depletion emerges around $p_{c1}$ and the integrated area of the \BAN peak collapse at $p_{c2}$.} 
\label{fig6}
\end{figure}

In summary, our two main Raman experimental observations in the superconducting state are originated from the impact of the pseudogap on superconductivity.  An uniform distribution of supercondcutivity around the whole Fermi surface is no more available for high-Tc cuprates. The pseudogap opens at  $p_{c1}$ and causes a progressive loss of coherent Bogoliubov quasiparticles around the antinodes and the emergence of two energy scales as we move to $p_{c2}$. In a such a way only fractions of coherent Fermi surface (for which coherent Bogoliubov quasiparticles have been detected) persist around the nodes. 
Below $p_{c2}$ superconductivity is mostly confined around the nodes.  



\section{A new relationship between the superconducting gap and $T_c$: a clue for increasing $T_c$? }
 
Once it has been shown that pseudogap is directly related to the loss of antinodal coherent Bogoliubov quasiparticles and the emergence of two energy scales in the superconducting state, it remains us to simulate the Raman spectra in the superconducting state by taking into account the impact of the pseudogap on superconductivity. In order to proceed we have considered a simple $d-$ wave function for the superconducting gap and a loss of coherent Bogoliubov quasiparticle spectral weight at the antinodes \cite{Blanc}. 

Within a Fermi liquid description we define the quasiparticle contribution to the Raman response by:

\begin{equation}\label{eq:response}
\chi^{''}_{\BAN,\BN}(\Omega)=
\frac{2\pi N_{F}}{\Omega}\left\langle \gamma^2_{\BAN,\BN}(\phi)\,(Z\Lambda(\phi))^2
\frac{\Delta(\phi)^2}{\sqrt{(\Omega)^2-4\Delta(\phi)^2}}\right\rangle_{FS}
\end{equation}
The angle $\phi$ is associated with momentum k on the Fermi surface. The gap function is described by $\Delta(\phi)=\cos 2\phi$. It vanishes at the nodal point $\Delta(\phi=45^0)=0$ while it is maximal at the antinodes $\Delta(\phi=0)=\Deltam$. Its amplitude increases as the doping level is decreasing as sketched in the top left panel of fig.7. 

The angular average over the Fermi surface is denoted $\langle(\cdots)\rangle_{FS}$. $N_{F}$ is the density of states at the Fermi level. We assume here that the density of states $N_F$ does not depend sensitively on doping level between $p=0.10$ and $0.20$. $\gamma_ {\BAN,\BN}$ are the Raman vertices which read $\gamma_{\BAN}(\phi)=\gamma^{0}_{\BAN}\cos 2\phi$ and
$\gamma_{\BN}(\phi)=\gamma^{0}_{\BN}\sin 2\phi$, respectively. $\Delta(\phi)^2/\sqrt{\Omega^2-4\Delta(\phi)^2}$ is the BCS coherence factor. 

The function $Z\Lambda(\phi)$ is the renormalized spectral weight of the Bogoliubov quasiparticles. $\Lambda(\phi)$ is a Fermi liquid parameter associated with the coupling of these quasiparticles to the electromagnetic field.

The effect of the pseudogap is to decrease the renormalized spectral weight $Z\Lambda(\phi)$ around the antinodes $\phi=0$ as the doping is reduced. To proceed in the simplest way, we consider a crenel-like shape for $Z\Lambda(\phi)$, centered at $\phi =45^o$ and define the angular extension $\phi_c$ for which $Z\Lambda(\phi)$ is significant . This corresponds to the fraction of coherent Fermi surface $\arc\equiv (45{^o}-\phi_{c})/(45{^o})$ inside which superconductivity settles. As the doping level is reduced \arc narrows and the crenel function shrinks.  The quasiparticle spectral weight (QPSW) remains significant around the nodes and decreases at the antinodes (see the left bottom panel of fig.7). \cite{Note0}

The angular dependence of the quasiparticle renormalization $\phi_c$ plays a key role in accounting for the disapearance of the \BAN peak and the emergence of the two energy scales. 

Let us first consider the \BAN geometry. The Raman vertex $\gamma_{\BAN}(\phi)$ is peaked at the antinode $\phi=0$ which dominates the \BAN response, resulting in a pair-breaking coherence peak at $\hbar\Omega_{\BAN}=2\Deltam$ due to the singularity of the BCS coherence factor. The weight of this peak is directly proportional to the antinodal quasiparticle renormalization $(\ZAN)^2=(Z\Lambda)^2(\phi=0)$. Hence, the fact that the \BAN coherence peak looses intensity at low doping (and even disappears altogether at low doping) is due to the decrease of $\ZAN$ as doping falls down. 
 
In the \BN geometry, the situation is more subtle because the Raman vertex is largest at the nodes, where the gap function (and hence the BCS coherence factor) vanishes. As a result, the energy of the coherence peak depends sensitively on the angular dependence of the quasiparticle renormalization $Z\Lambda(\phi)$. If the latter is approximately constant along the Fermi surface, then the energy of the \BN peak is determined solely by the angular extension of the Raman vertex $\gamma_{\BN}(\phi)$. In contrast, the angular extension $\phi_c$ around the node is smaller than the intrinsic width of the Raman vertex $\gamma_{\BN}(\phi)$ (see left bottom panel of fig.7). Then, it is $\phi_c$ itself which controls the position of the \BN peak: $\hbar\Omega_{\BN}=2\Delta(\phi_c)$. This explains the origin of the differentiation between the two energy scales in the underdoped regime: while the \BAN coherence peak increases in energy with falling doping, $\Delta(\phi_c)$ decreases because of the rapid contraction of the coherent fraction $\arc$, leading to the decrease of the \BN peak energy and therefore the opposite doping dependence of the two scales as illustrated in the right pannel of fig.7. We can also notice that because the QPSW is still significant around the nodes, the intensity of the \BN peak remains detectable while the \BAN peak intensity strongly decreases and disapears for moderately underdoping.

\begin{figure}[!ht]
\begin{center}
\includegraphics[width=11cm]{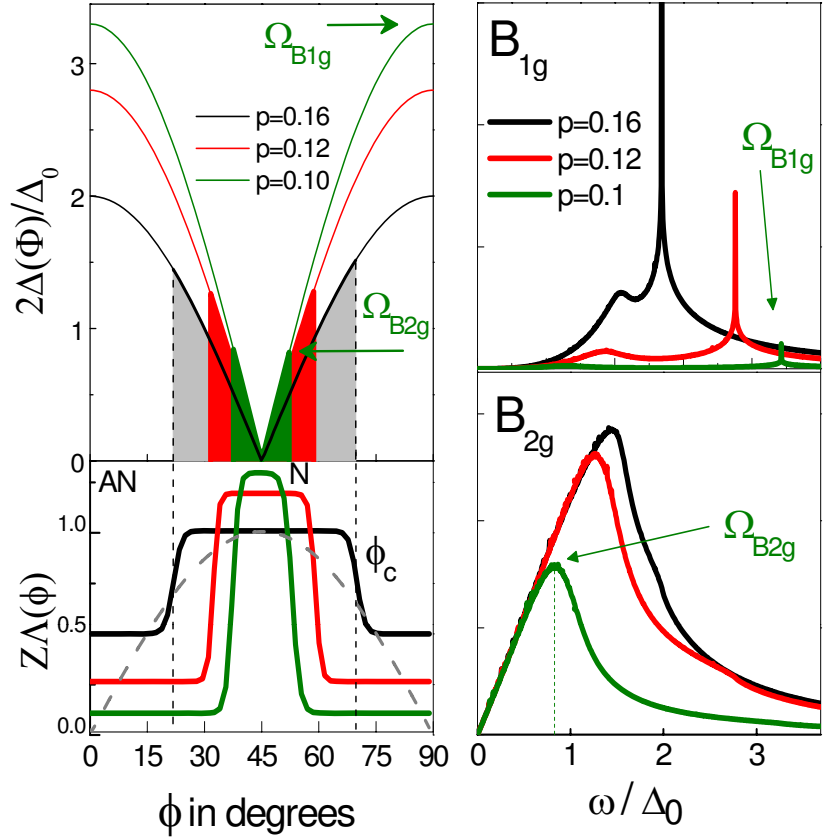}
\end{center}\vspace{-7mm}
\caption{Left panel: at the top, doping evolution of a single d-wave superconducting gap. The shaded area correspond to the fraction of coherent Fermi surface, \arc, around the nodes. At the bottom: the doping dependence of the quasiparticle spectral weight (QPSW). $\phi_c$ angle delimits the extension of \arc.  When $\phi_c$ is smaller than the \BN vertex width (gray dashed line) $\phi_c$ itself controls the position of the \BN peak. Right panel: Calculated Raman spectra in \BAN and \BN for several doping levels.} 
\label{fig7}
\end{figure}

Interestingly, we note that linearizing the gap function in the nodal region (see top left panel of fig.7) is a reasonable approximation, leading to the relation $\hbar\OmN=45{^o}\arc\vD\propto k_{B}T_c$ which links the nodal (\BN) energy scale (proportional to $T_c$), the nodal velocity and the coherent fraction \cite{Note1}. Since for a simple d-wave gap : $\vD\propto\Deltam$,  the relation between the critical temperature (or $\OmN$) and the coherent fraction reads: $k_BT_c \propto \arc\Deltam$. This relation carries a simple physical meaning, namely that it is the suppressed coherence of the quasiparticles that sets the value of $T_c$, while $\Deltam$ increases with falling doping. This relation differs from the standard BCS theory. Crucially, $T_c$ in cuprates depends on the gap $\Deltam$  but also a prefactor, \arc which is doping dependent.

Our interpretation reconciles the distinct doping dependence of the two energy scales with the thermal conductivity measurements of underdoped cuprates.  Quasiparticle thermal conductivity measurements interpreted within the clean limit and a Fermi velocity $v_F$ almost constant \cite{Note2} show that $K_{0}/T\propto\frac{v_F}{\vD}$  decreases with falling doping and  $\vD\propto\Deltam$ \cite{Sutherland,Hawthorn}. 

It is also in agreement with previous observations on Giaver and Andreev Saint-James (ASJ) tunneling experiments which pointed out the existence of two distinct energy scales in superconducting state of underdoped cuprates \cite{Deutscher}. The high energy scale was assigned to the single particle exictation energy. This is the energy of the first excited state required to break a Cooper pair in Giaver tunneling experiment \cite{Giaver}. This corresponds to the Raman \BAN scale associated to the pair breaking peak energy. The low energy scale was assigned to the energy range over which Cooper pairs can flow in the ASJ tunnelling. It is directly related to the Raman \BN scale since, this last one, is controlled by \arc around the nodes where supercurrents flow. 

Our experimental observations and simulations show that $T_c$ is limited by \arc such as $k_BT_c \propto \arc\Deltam$. In parallel way, such a relationship can be also achieved by considering the Uemura relation (valid in the underdoped regime) and Homes' law  \cite{Uemura,Homes}. Indeed, $\rho_{S}\propto T_{c}$ and $\rho_{S}\propto \sigma_{dc}\Deltam$ (valid in the dirty limit) lead to $T_{c}\propto \sigma_{dc}\Deltam$. $\rho_{S}$ and $\sigma_{dc}$ are respectively the superfluid density and dc-conductivity.  $T_{c}$ is thus driven by the maximum amplitude of the superconducting gap and the physical quantity which controles the current flows namely $\sigma_{dc}$ or \arc. Our prediction is therefore that cuprates with the largest $\sigma_{dc}$ above $T_c$ close to the optimal doping level will give the highest $T_c$. 
This explains why $T_c$ can be reduced in Zn and Ni-substituted optimally doped Y-123 compounds although $\Deltam$ remains unchanged \cite{Gallais2,leTacon2}. The reason is that $\sigma_{dc}$ decreases by Zn and Ni substitutions while preserving the carrier concentration.

In conclusion, our experimental findings reveal that the superconducting state is disturbed by the emergence of the pseudogap. Contrary to the conventional superconductors for which all the electronic states near the Fermi level are involved in the superconducting state, only a part of these electronic states is involved in superconductivity of underdoped cuprates. This disturbance of the Fermi surface is caused by the emergence of the pseudogap which develops along the antinodes. This gives rises to three regions inside the superconducting states delimited by $p_{c1}$ and $p_{c2}$ which signal respectively the begining and the end of the disturbance of the Fermi surface at the antinodes. 
Between $p_{c1}$ and $p_{c2}$, Fermi surface transforms into fractions of coherent Fermi surface centered around the nodes. The coherent Bogoliubov quasiparticles are distroyed around the antinodes. This k-space dichotomy leads to two energy scales in the superconducting state. The \BAN energy scale is related to the maximum amplitude of the superconducting gap at the antinodes and the \BN one to the non vanishing amplitude of the superconducting gap at the ends of the coherent fractions of the Fermi surface.  

Although the superconducting gap increases with falling doping level, the loss of well defined Bogoliubov quasiparticles in the antinodal region locks $T_{c}$ and leads to its decrease. Preserve available coherent Bogoliubov quasiparticules at the antinodes for superconducting current flow is from our point of view the crucial point for increasing $T_c$.

\par

\subsection{Acknowledgments}

We thanks A. Georges, G. Kotliar, A. Millis, J. Tallon, G. Blumberg, M. Civelli, S. Sakai, C. Proust, 
M. Le Tacon, S. Blanc, S. Benhabib and Z.Tesanovic for their participation to very fruitful discussions. 

\section{References}

\end{document}